# Scientists are Working Overtime and at the Weekends: Comparison of Publication Downloading from Copyrighted and Pirated Platforms

Yu Geng[a], Ren-Meng Cao[a], Xiao-Pu Han[b,c], Wen-Can Tian[a], Guang-Yao Zhang[a], Xian-Wen Wang[a,*][1]

a. WISE Lab, Faculty of Humanities and Social Sciences, Dalian University of Technology, Dalian 116085, China

b. Alibaba Research Center for Complexity Sciences, Hangzhou Normal University, Hangzhou 311121, China

c. Institute of Information Economy and Alibaba Business School, Hangzhou Normal University, Hangzhou 311121, China

**Abstract:** In this study, we track and analyze publication downloads from both copyrighted and pirated platforms to reconstruct scientists' activity patterns from a holistic perspective. Scientists around the world are working overtime, but scientists in different countries have different working patterns. Scientists' preferences for different platforms are influenced by a variety of factors such as working times and workplace arrangements. There are variations by country in terms of whether scientists prefer to work overtime at night, at the weekend, or both at night and on the weekend. When scientists are working overtime, they prefer to use Sci-Hub rather than copyrighted platforms to access scholarly publications This may be because of the transition in their working scenarios as they move from the office to home outside of work hours.

**Keywords:** Working overtime; work–life balance; Sci-Hub; off-campus access; pirated papers; copyrighted papers

## 1. Introduction

### 1.1 "Publish or perish" and the pressure on scientists

Scientists in the scientific community are frequently under pressure to publish early in their academic careers, and the culture of "publish or perish" not only encourages academic misconduct (Qiu, 2010) but also creates an imbalance between the work and personal lives of scientists, resulting in a slew of issues. Many authors have looked into the work habits of researchers with this in mind.

Work is the primary source of revenue; moreover, it is directly tied to one's social stratum (Schieman et al., 2009). However, as the pressure of work increases, the degree to which it interferes with family life also increases, and family life can be sacrificed to work when people take work problems home to solve (Frone, 2003; Jacobs & Gerson, 2004; Kelloway et al., 1999). Scientists' working lives are a good example of this. To keep focused on their academic problems, scientists give up a great deal of their free time. Some previous studies looked at scientists' work habits and discovered that there are unwritten overtime norms in academia and that many scientists work deep into the night and on weekends (Fox et al., 2011; Wang et al., 2013).

---

[1]* Corresponding author: xianwenwang@dlut.edu.cn

Working overtime may be one method for scientists to cope with the "publish or perish" tendency. Cabanac et al collected and examined the publication histories for articles published in *JASIST* to research the work-life balance issues among authors and editors. It is found that 11% of manuscript-related events happened during the weekends(Cabanac & Hartley, 2013). This finding is consistent with Wang et al.'s(Wang et al., 2013) results about overworking scientists.

Scientists' working time patterns are also influenced by their work environment and culture, which are represented at the macro level and vary widely from one nation to another. We can even estimate a scientist's working hours depending on the country in which they reside. China, for example, is typical of a country with a culture that encourages diligence, where researchers' performance and advancement are tied to their scientific findings; thus, Chinese researchers are under extreme pressure to "publish or perish." In contrast, people are less likely to work outside of working hours in countries with congenial working conditions, minimal work pressure, and unions that protect academic researchers' rights (Barnett et al., 2019; Bentley & Kyvik, 2012; Huang, 2018; Hvistendahl, 2013).

## 1.2 The battle between copyrighted and pirated papers

Scientists' submission or peer review time records (Barnett et al., 2019), records of the times at which scientists download papers (Wang et al., 2013), and lighting data records are all examples of typical tracking methods for studying scientists' work times (Wang & Ma, 2017). In this study, we are interested in the situation of scientists downloading papers from pirated platforms besides t copyrighted publishers.

There is a long history of conflict between copyrighted and pirated copies of academic articles. The Sci-Hub website was founded in 2011 by Alexandra Elbakyan. Sci-Hub provides free access to around 88 million publications, blatantly infringing on the copyrights of major publishers such as Elsevier, Springer Nature, and Wiley-Blackwell. Sci-Hub has faced various challenges from copyright holders and US legislation since its inception (Bezerra & Sanches, 2018; Murphy, 2016; Schiermeier, 2017). It is, however, growing in popularity in the academic community. Sci-Hub has users in every country in the world, and a piece in *Science* only slightly exaggerates the site's popularity when it says that "Everyone is downloading pirated publications" (Bohannon, 2016).

The way scientific discoveries are published and communicated has changed as a result of digitization and the Internet, making knowledge more accessible and widely disseminated. However, for-profit publishers have monopolized knowledge, and expensive paywalls have stifled scientific progress (Siler, 2017). Some studies claim that monetizing scientific knowledge and putting it behind paywalls violates the concept of fairness and hurts the public interest and suggest that open access piracy is, therefore, a justified act of civil disobedience from an ethical standpoint (James,

2020). Some argue that the resistance movement against academic capitalism is a new type of "digital socialism," noting that in-depth interviews with researchers in Central and Eastern European nations have helped clarify the ideas of "academic capitalism" and "academic imperialism" (Łuczaj & Holy-Łuczaj, 2020). The breadth of material covered by Sci-Hub as well as its ease of use contribute to its popularity. As of March 2017, Sci-Hub's database contained 68.9% of Crossref-registered scholarly publications and 85.1% of articles published in fee-access journals (Himmelstein et al., 2018). Free, accessible, and quick access to literature is preferred by users (Chen, 2018).

Of course, there is vociferous opposition to Sci-Hub, with for-profit publishers claiming that Sci-Hub violates their rights by collecting their publications without permission. When researchers use Sci-Hub to access papers, the download statistics for the literature are not properly recorded, which has several potential consequences, including authors not being able to benefit from download statistics and libraries being unable to track the use of the journals they provide. Even though the cost of reproducing and disseminating digital literature is negligible, the people who work for journals, such as editors, proofreaders, illustrators, and science communicators, are professionals who must be paid and who add value, resulting in the high cost of digital publications (McNutt, 2016).

Unlike previous studies using either copyrighted or pirated platforms, we analyze data from both kinds of platforms, so we can gain a full picture of scientists' working patterns by integrating copyrighted and pirated platforms.

In this study, we focus on the following research questions:

(1) What are scientists' typical working patterns?

(2) How do scientists' behaviors differ when they access articles from copyrighted or pirated platforms?

(3) What motivates scientists to download more pirated papers at night and weekends than during working hours?

## 2. Data and Methods

Our dataset for this study is made up of two parts.

The first set of data was taken from Springer's real-time platform (Wang et al., 2013), which collects data in real time for 24 hours and can track users' download activity as well as their location. We collected data from April 8, 2012, through April 14, 2012. Each country and region's time zones were enumerated, and the time in GMT was converted to the local time in each country and region. We ended up with a total of 1,048,575 data points.

The second source of data was the Sci-Hub platform server data, which were shared via the @scihub account on the Twitter social media platform (https://twitter.com/Sci_Hub, which Twitter has since disabled). The server that hosts the Sci-Hub website is in Russia (Bohannon, 2016), and the server's time zone is near GMT+3 (*Archive Ancient Sci-Hub Download Logs (2011-2013) · Greenelab/Scihub@61ba4a0*, n.d.). We converted the time to the local time in each country and region after converting the time zones.

The Sci-Hub dataset covers the period from January 1, 2017 to December 31, 2017, and contains 140 million records. We examined the 2017 resident population data for a number of regions and found some regions with very small populations but huge download volumes. This part of the data seems to be outliers. As Sci-Hub allows large-scale downloading of papers using scripts/crawlers/robotic programs and other means, there is also the possibility of using IP proxies to download papers. We used scripts to identify the data of machine-downloading behavior in the dataset and removed data with incomplete fields, resulting in 74,742,113 valid data points.

## 3. Results

Because the total number of downloads from the same country in the Sci-Hub dataset and Springer dataset vary widely and cannot be compared directly in order of magnitude, the proportion of downloads in different time dimensions must be transformed into a percentage form. China, India, Germany, and the United States, among others, have the highest download rankings in our dataset. Based on the number of downloads, download volume, and geographic distribution, we chose some typical countries as representative of North America (the US, Mexico), South America (Brazil), Asia (China, Japan), Europe (Germany, the UK), and Africa (South Africa). Table 1 displays the number of publications downloaded from the Sci-Hub and Springer platforms for each of the dataset's eight typical countries.

Table 1 Number of Sci-Hub downloads and number of Springer downloads in the sample countries

| Country | Sci-Hub download_number （2017.01~2017.12.31） | Springer download number （2012.04.08~2012.04.14） |
| :---: | :---: | :---: |
| China | 11,316,634 | 125,958 |
| Brazil | 5,442,464 | 17,895 |
| United States | 4,082,952 | 324,609 |
| Germany | 2,167,967 | 137,339 |
| Mexico | 1,793,349 | 9,105 |
| Japan | 796,035 | 34,021 |
| South Africa | 195,983 | 4,581 |
| United Kingdom | 1,544,080 | 40,031 |

### 3.1 Working patterns

The time series overlay analysis method was used to calculate the number of downloads for each moment in the dataset from 0:00 to 24:00, and then the percentage was calculated by dividing the number of downloads by the total number of downloads for each hour. There was a significant difference in absolute numbers between Springer and Sci-Hub, and using the percentage format helps to visually quantify the relative differences between the two download sources.

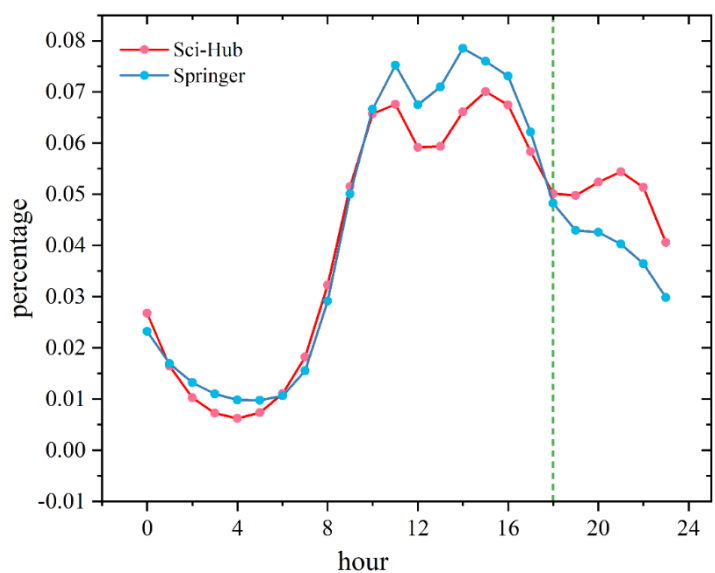

Fig. 1 24-hour activity distribution of the dataset

Fig. 1 shows that downloads are at their lowest point of the day around 04:00, when most people are sleeping and only few are working. All downloads achieve their first peak of the day around 10:00–11:00, when work efficiency is at its peak; 12:00–13:00, which is the time for lunch and maybe a rest, ushers in a fallback period. The second download peak of the day occurs between 14:00 and 15:00, which is the most productive working hour in the afternoon session.

In summary, the waveforms of Sci-Hub are extremely comparable to those of Springer from 00:00 to 18:00; then the waveforms begin to exhibit substantial variances between 18:00 and 24:00; 18:00 marks a turning point, separating scientists' work and personal lives. After 18:00, the download patterns for Sci-Hub and Springer begin to differ, owing to the fact that most individuals leave work around 18:00. After work, Springer's download distribution shows a parabolic decrease, whereas the download rate for Sci-Hub continues to grow, hitting the day's third download peak at 21:00. We believe the shift after 18:00 is correlated to a shift in people's personal space. In other words, people transfer from their offices to their homes but continue to work, and because they do not have easy access to official academic platforms in their homes, they prefer to get papers from pirated academic platforms. As a result, the download curves for Sci-Hub and Springer diverge significantly.

### 3.2 Differences across countries and regions

In order to compare scientists' preferences for different platforms, the eight typical nations and regions that were picked from the dataset would be analysed.

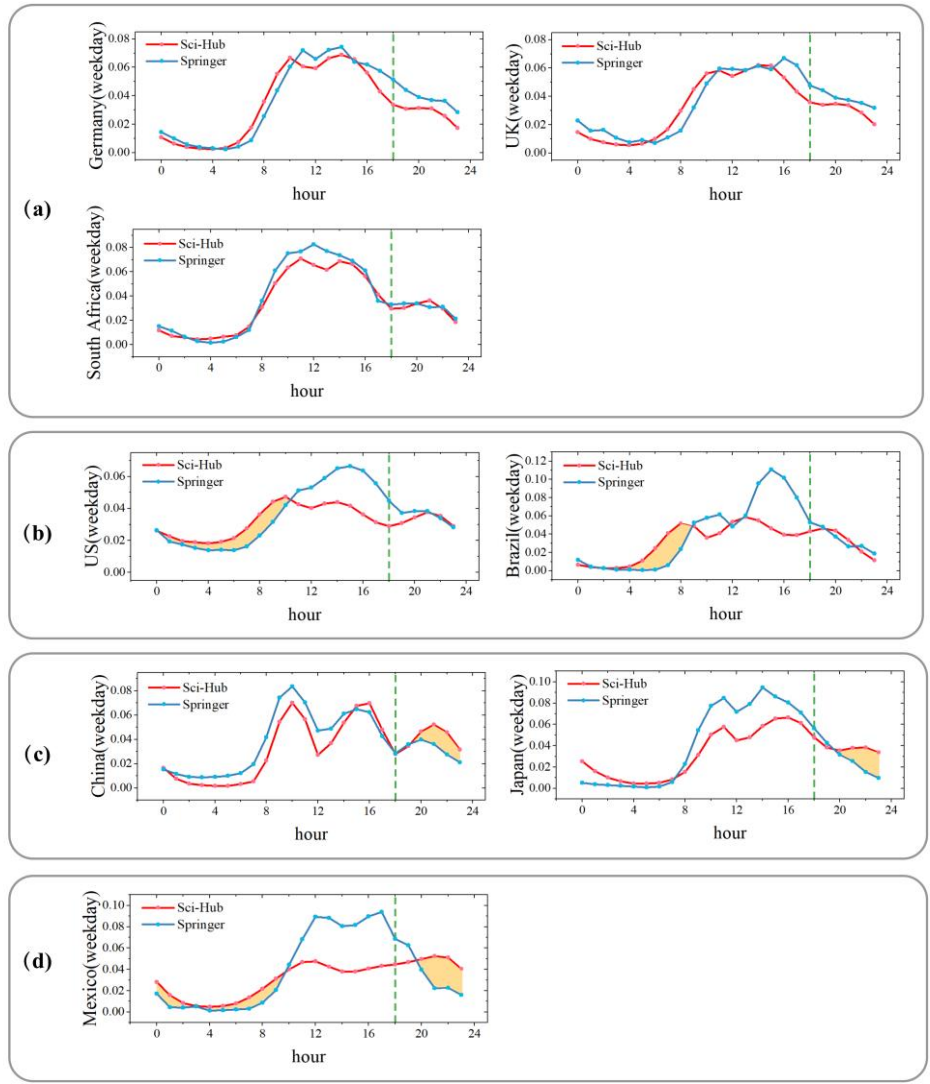

Fig. 2 The download distribution curves are divided into four groups, group (a) represents the three countries with similar working patterns and less overtime; group (b) represents the two countries focus on working overtime during early morning; group (c) represents the two countries focus on working overtime at night; group(d) represents the country's working pattern of overtime in the evening and early morning.

As can be seen, the eight countries are classed based on whether they work overtime during off-hours on weekdays(Fig. 2). Fig. 2(a) shows that Scientists in Germany, the United Kingdom, and South Africa work standard office hours, and the number of articles downloaded after 18:00 is falling, indicating that the numbers of scientists in those nations who work late are low.

It can be observed that during the early morning, scientists in the United States focus on working overtime between 00:00 and 10:00; and scientists in Mexico focus on working overtime between 00:00 and 9:00 (Fig. 2(b)).

Typical working overtime at night can be seen in Fig. 2(c). Scientists in China and focus on working overtime between 20:00 and 21:00; scientists in Japan focus on working overtime between 21:00 and 23:00.

Similar behavior was observed in Fig. 2(d). Scientists in Mexico focus on working overtime between 00:00 and 10:00 during the early morning, and between 20:00 and 23:00 at night.

The download curve has considerable geographical peculiarities, according to our findings. Scientists in Asia and North America are accustomed to working overtime in the evenings, whereas scientists in South America, Europe, and Africa distinguish work from life by being more attentive during work hours and working less overtime during off-hours.

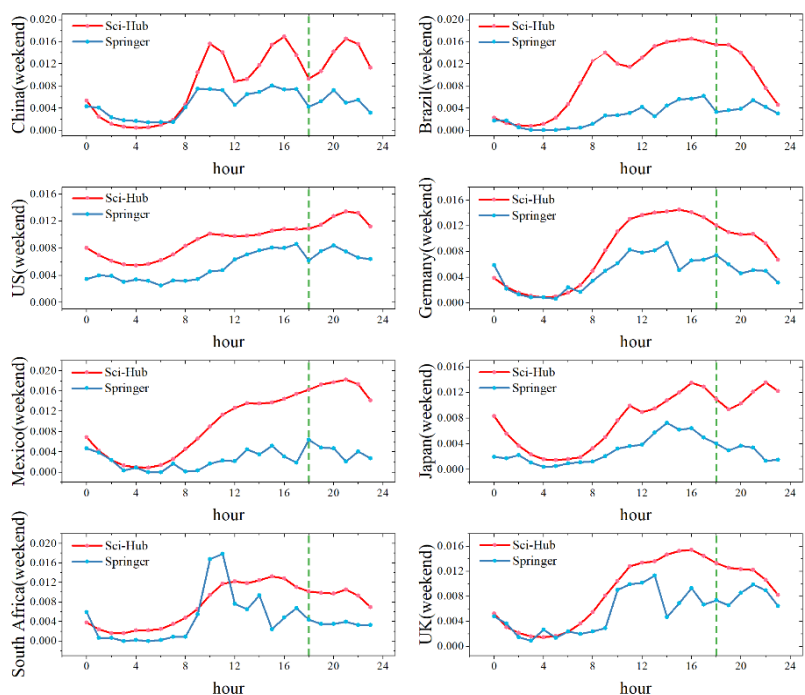

Fig. 3 Download distribution of eight representative countries on weekends

as can be seen in Fig. 3, the weekend distribution of 24-hour download percentages for each country is notably different from the weekday distribution. We observe that China and Japan in Asia, the United States and Mexico in North America, and the United Kingdom in Europe work overtime on weekend evenings, implying that these five countries maintain their work habits and overtime habits throughout the weekend. Brazil, Germany, and South Africa, among the other three countries, all work overtime on weekends, but the intensity of work on weekend evenings decreases steadily.

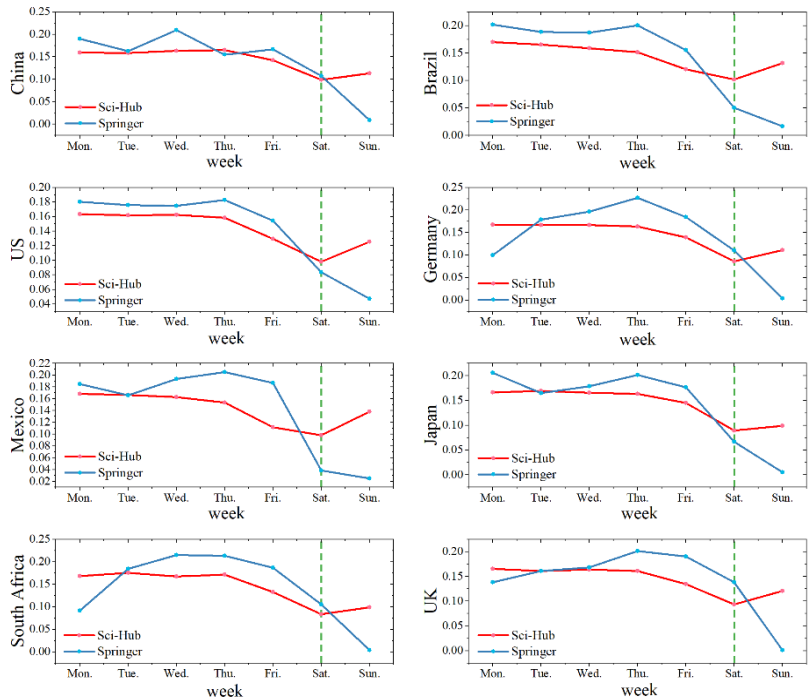

Fig. 4 Download distribution of eight representative countries in one week

We counted the distribution of downloads over the week in eight countries to learn more about scientists' work in a week. In these figures, as shown in Figure 4, Saturday becomes the pivotal day of the week, as people in most countries and areas of the world, with the exception of Muslim countries, take the weekend off.

The Springer downloads for the week, particularly the steep drop in the percentage of downloads from Saturday to Sunday, might lead one to the conclusion that "scientists take the weekends off." However, this is not the case, and the reality can be seen in Sci-Hub's download curve.

We noticed that on Saturday, the download trend of Sci-Hub reversed, and instead of declining from Saturday to Sunday, the download rate of Sci-Hub climbed, and the download waveform of Sci-Hub looked like a tail, indicating that scientists took a break on Saturday before returning to their research on Sunday.

### 3.3 Transfer of work scenarios

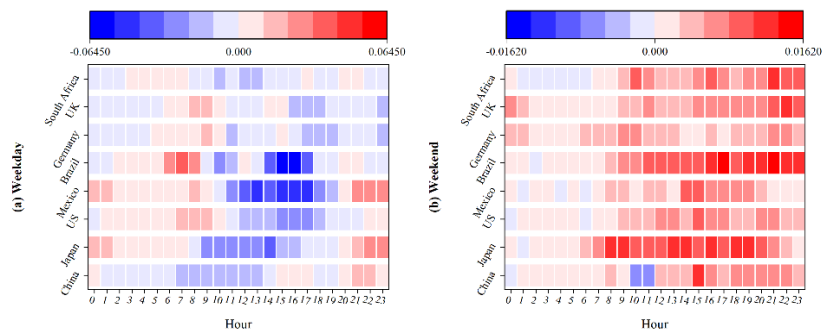

Fig. 5 Heat map of the distribution of 24-hour download differences for eight

representative countries

We utilize a heat map to observe the distribution of differences between periods in order to finely differentiate them. We define the heat map's value as *P(t)*:

$$P(t) = P_{Sci-Hub}(t) - P_{Springer}(t)$$

*P*<sub>Sci-Hub</sub>(*t*) is the download ratio of Sci-Hub at moment *t*, *P*<sub>Springer</sub>(*t*) is the download ratio of Springer at moment *t*. *P(t)* is the difference between the download ratio of Sci-Hub minus the download ratio of Springer at moment *t*. If *P(t)* > 0, the download ratio of Sci-Hub at moment *t* is higher than that of Springer. If *P(t)* = 0, the download ratio of Sci-Hub at moment *t* is equal to that of Springer. If *P(t)* < 0, the download ratio of Sci-Hub at moment *t* is lower than that of Springer.

The *P(t)* indicator can only show the difference in size between Sci-Hub and Springer, not the number of downloads. As a result, although the red section of the heat map may not have a high download ratio, Sci-Hub's download ratio must be larger than Springer's, and we must study it in conjunction with the findings in Section 3.2. The question we need to investigate is whether scientists use Sci-Hub or Springer to download papers during their off-hours.

China, Japan, the United States, and Mexico are four examples of countries where scientists work overtime on weeknights. Fig. 5(a) shows that after 18:00, scientists in China, Japan, and Mexico primarily use Sci-Hub to obtain pirated publications, whereas scientists in the United States use both sites.

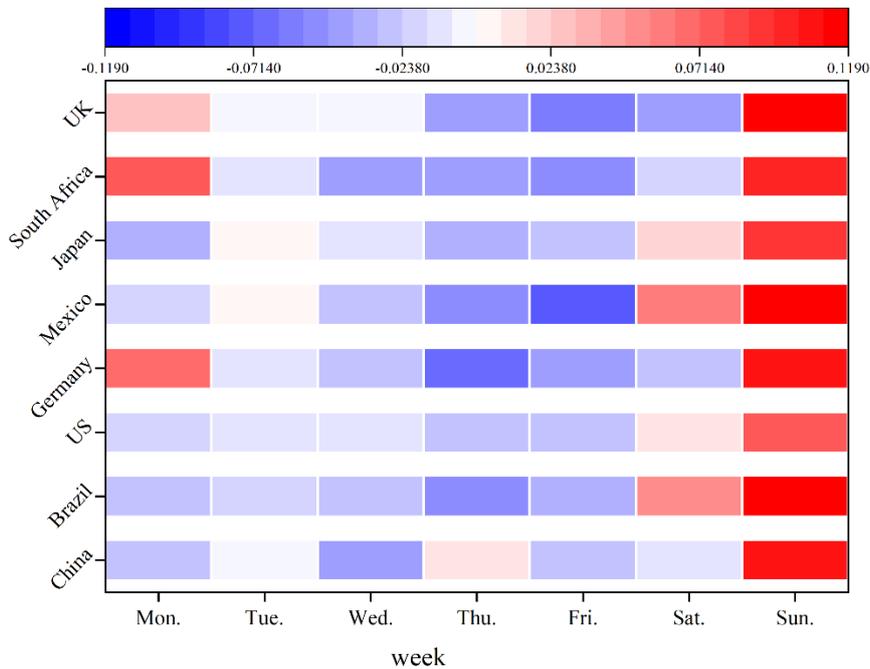

Fig. 6 Heat map of the distribution of week download differences for eight representative countries

Fig. 6 shows that on Monday, only Germany, South Africa, and the United

Kingdom use Sci-Hub to download publications, but from Tuesday to Friday, practically all nations and regions use Springer. On Saturdays, scientists in Brazil, the United States, Mexico, and Japan prefer to use Sci-Hub, but toward the end of the week, scientists in practically every country make heavy use of Sci-Hub to download papers.

The number of scientists working overtime in each country and region, as well as the frequency with which Sci-Hub is used, increases from Saturday to Sunday. During weekday off-hours and weekend breaks, scientists are more inclined to stay at home rather than go to the office. It is obvious that scientists' work schedules have shifted from work to break time, and this shift in work schedules has resulted in a distinction between scientists obtaining publications from copyrighted and pirated sites.

## 4. Discussion

### 4.1 The phenomenon of scientists working overtime

Prior work has documented the phenomenon of scientists working overtime, Xianwen Wang, for example, reports that scientists essentially do not take weekends off, and scientists always focus on working overtime even on weekdays. However, these studies have either been sort-term studies or only have focused on the Springer platform. In this study we tested the extent to with an extended twelve-month download history of the Sci-Hub, and a comparison was made based on data from the Springer study by Xianwen Wang.

We found that scientists download more pirated papers in their rest time, and scientists work more intensely on downloading pirated papers during their breaks than they do on downloading copyrighted papers. These findings extend those of Xianwen Wang, confirming that scientists work more heavily during off-hours and weekends than previous studies have observed, the paper download record of the Springer platform is just the tip of the iceberg of the scientists' hard work.

### 4.2 Time and scenario analysis of scientists downloading pirated literature

Scientists use Sci-Hub more frequently to access literature on weekday nights and non-working days, according to the study, a pattern that caught our attention. On weekday nights, scientists in countries such as China, Japan, and Mexico visit Sci-Hub more frequently; on weekday nights, scientists in countries such as the US, Brazil, and South Africa use Sci-Hub at rates similar to those at which they use Springer; scientists in countries such as the UK and Germany use Sci-Hub at rates that are slightly lower than those at which they use Springer. Although significantly lower than that of Springer, the percentage of scientists accessing Sci-Hub on weekday evenings in countries like the UK and Germany is, nevertheless, not negligible. Furthermore, the usage of Sci-Hub on non-working days in practically all countries greatly outnumbers that of Springer.

According to the findings, scientists' frequent use of pirated literature is not an unusual incident, but rather a common practice. The use of Sci-Hub by scientists to download pirated literature is linked to factors such as overtime hours and workplace arrangements. In the 24 hours of the day, scientists may change workplaces several times, alternating between offices, restaurants, houses, and other locations as they transition from work hours to non-work hours on weekdays, and the situation may also be different on weekends. Scientists are most likely to be in the office during weekday office hours but are more likely to be at home after dinner, and Sci-Hub has a larger percentage of downloads than Springer during this time. Similarly, scientists are more likely to be at home on weekends than in the office, and Sci-Hub clearly outperforms Springer in terms of weekend downloads.

### 4.3 Why do scientists download pirated literature?

What is the timing and situation of scientists downloading papers from Sci-Hub? One possible factor is the "paywall" mechanism that publishers have implemented to prevent people from accessing the material freely. Publishers construct monopolies through copyright rules and charge hefty royalties through paywalls, which not only makes it difficult for scientists to access papers through legal channels but also deepens economic inequalities in academia and stifles intellectual exchange (Iván Farías & Flor, 2016). When scientists work from home, they use off-campus access services to access information resources. The off-campus access services, however, do not perform as effectively as they should, and there are numerous barriers to their use. Researchers must go through a variety of authentication protocols to gain off-campus access, including using off-campus VPNs, proxy server logins, website registration logins, or logging in to access resources using the subscribing organization's username and password, which makes it inconvenient to access copyrighted content through off-campus access and has encouraged some to turn to illegal piracy sites (Hurst & Schira, 2019; Moore, 2020).

Although numerous hurdles stand in the path of scientists' scientific studies, publishers' paywalls, inconvenient authentication methods, and other factors surely consume and waste a great deal of scientists' time, perhaps increasing their burden and lengthening their working hours.

Nobody enjoys being inconvenienced. Tury et al. studied remote information access behavior by surveying 649 students from 81 countries. They discovered that the most essential elements were accessibility and speed of access to information, as well as familiarity with the source and concluded that these characteristics were more relevant than issues of quality, reliability, or comprehensiveness (Tury et al., 2015).

As a result, making copyrighted literature more accessible to researchers would be great beneficial to science. On the one hand, open access is the best way to enable knowledge sharing. In the case of open-access literature, researchers can access scientific literature at any time and from any location. Academic publishers, institutional libraries, and institutional IT services, on the other hand, may collaborate to create a better and easier environment for researchers to access the literature and remove barriers to legitimate use, such as providing seamless, privacy-protected, one-click subscription content so that scientists can access articles legitimately whenever and wherever they want (Felts et al., 2020; Moore, 2020).

In conclusion, this is the first study to our knowledge to investigate the phenomenon of scientists downloading pirated papers during overtime. Our results provide compelling evidence for long-term overtime among scientists and suggest that this approach appears to be effective in proving that scientists do more unknown hard work in their break time. Future work should therefore include follow-up work designed to examine whether the content and topics of papers downloaded by scientists during work time are consistent with those downloaded during break time, and whether scientists study these papers due to work responsibilities or due to interest.